\documentclass[twocolumn,aps,prl,showpacs]{revtex4}
\usepackage{graphicx}
\newcommand{\be}{\begin{equation}}
\newcommand{\ee}{\end{equation}}
\newcommand{\ben}{\begin{eqnarray}}
\newcommand{\een}{\end{eqnarray}}
\begin{document}

\title{Solitons in Poly(oxyethylene)}
\author{D. Bazeia$\,{^a}$, M.Z. Hernandes$\,{^b}$, and A.M. Simas$\,{^c}$}

\affiliation{$^a$Departamento de F\'\i sica, Universidade Federal
da Para\'\i ba, 58051-970 Jo\~ao Pessoa,
Para\'\i ba, Brazil\\$^b$Departamento de Farm\'acia, Universidade Federal
de Pernambuco, 50740-521 Recife, Pernambuco, Brazil\\$^c$Departamento de
Qu\'\i mica Fundamental, Universidade Federal de Pernambuco,
50670-901 Recife, Pernambuco, Brazil}
\date{\today}
\begin{abstract} 
We introduce a soliton model to describe conformational structure
of poly(oxyethylene) in aqueous solution. The model is based
on former investigations for twistons in polyethylene, and the soliton
solution corresponds to stable excitation that locally transforms
the {\it gauche} conformation of the O-CH$_2$-CH$_2$-O dihedral angle from the
positive (negative) configuration to its negative (positive) partner.
The conformational deformation in the polymer main chain is mediated
by the unstable {\it cis} conformation which changes the physical profile of
the chain, contributing to trigger important properties of the polymer.
For instance, it may activate the process for the capture of metallic ions
in aqueous solution and it may also enlarge solubility of the polymer in
water. The soliton solution is used to guide a quantum chemistry investigation
that examine, and characterize, a chain containing 40 monomeric units, tracing
quantitative correlations with the soliton model.
\end{abstract}
\maketitle

Solitons appear under the direct influence of nonlinearity, and may play
important role in several branches of nonlinear science. In particular, they
may spring as finite energy excitations in nonlinear polymeric chains. A
classic example is polyacethylene (PA): in this chain, single and
double bonds alternate, generating Peierls instability that gives rise to
solitons, as accounted for in the Su-Schrieffer-Heeger
model \cite{ssh79}; see Ref.~{\cite{hks88}} for further details.
As one knows, the PA chain admits an interesting field theoretical
formulation in terms of bosonic and fermionic fields \cite{jsc81,jac95}.
Another example of a nonlinear
polymeric chain that supports solitonic excitations is polyethylene (PE):
in this case the chain only contains single sigma bonds, which activate
degrees of freedom of the torsional type, giving rise to twiston excitations
in the polymeric chain \cite{mbo78}. The presence of twistons in the PE chain
was further explored in Refs.~{\cite{80,zco93}}, and more recently in
Ref.~{\cite{we}} one has shown that these twistons can be seen as topological
solitons that appear in a specific model of two real scalar fields.

The work of Ref.~{\cite{we}} follows the investigations on topological
solitons iniciated in \cite{95}, and further explored more recently in
{\cite{bnr,bbr00,ilg}}. The main idea is that the presence of topological
solitons in a polymeric chain may directly contribute to the conformational
characteristics of the polymer, a fact important to the understanding of
physical, chemical and biological properties of the substance. A singular
example is the poly(oxyethylene) (POE), which has long been the subject of
great attention due to its various biotechnical and biomedical
applications \cite{har92}. In this letter we develop the idea that the POE
chain is able to support topological solitons, and that these solitons
contribute significantly to the undertanding of important features of this
polymer.

We start with the PE chain, which is a very long chain made of CH$_2$ units.
It may be seen as the limit for large $n$ of the chain family
[(CH$_2$)$_n$-O-]$_m$. In this family the second member ($n=2$) is the POE,
which is formed by repetitions of the chain unit CH$_2$-CH$_2$-O-.
The POE is a very important polymer in this family, because the peculiar
array in the chain provides to this polymer some very impressive features,
among them the remarkable ability to entrap metallic ions in aqueous solution
and its solubility in water to almost any extent. Besides, one knows that
ethylene oxide polymers are amongst the most commonly used substances in
pharmaceutical and other industrial formulations
\cite{har92,jpc93,jacs1995,jpca97,tas96,tas99,peg01,mat02}.

Molecular dynamics simulations \cite{smi96,tas96,prl2000} on the
conformational structure of the POE and 1,2-dimethoxyethane (DME =
CH$_3$-O-CH$_2$-CH$_2$-O-CH$_3$, which is correlated with the POE's dimer)
have shown that the polymeric chain may present a very rich family of
conformational possibilities, with preference for $tgt$ conformers in
condensed media -- we are using standard notation, where `$t$', `$c$' and `$g$'
represent {\it trans} ($\pi$), {\it cis} ($0$) and {\it gauche}
($\pi/3$) arrangements, respectively.  The $tgt$ conformer was
also observed in POE chains used as hidrophilic moiety of nonionic
surfactants, in aqueous solutions, through Raman spectroscopy \cite
{poe-raman}. The stabilization of the $tgt$ conformer in the liquid or solid
state may be understood in terms of polar intermolecular interactions,
because this conformer has an appreciable dipole moment while the $ttt$
conformer (all {\it trans} conformations) has zero dipole moment \cite
{jacs1995}. This preference for $tgt$ conformation is also predicted for DME
by experimental NMR \cite{dme-nmr} and infrared \cite{dme-ir} spectroscopic
studies. The water affinity of POE is directly related to a structural
similarity between the POE chain and liquid water, which is not found for
other polyethers \cite{kf81,ngm01}: when substantially hydrated, the POE
chain assumes local conformations mostly with the $tgt$ arrangement; in
this case there is almost perfect match between the distance of the
oxygen atoms in the structure of liquid water and the
distance between the nearest neighbor oxygen atoms in the polymer
chain. Thus, the POE chain may be easily incorporated into the structure
of liquid water, with the ether oxygen atoms replacing some of the water
molecules, with the hydrophobic ethylene groups being encaged in voids
in between water molecules. There are other conformational possibilities
in the POE chain; for instance, the $tgg$ conformer is considered
the second most populated in DME, also based on spectroscopic results
both for DME\ and POE, probably because of the presence of an important
intramolecular 1,5-CH$\cdots$O interaction that stabilizes this particular
conformation. The nomenclature $ttt$, $tgt$, and so forth, that we are using
is valid to represent the three most important dihedral angles
in the polymer main chain, namely O-CH$_2$-CH$_2$-O, CH$_2$-CH$_2$-O-CH$_2$,
and CH$_2$-O-CH$_2$-CH$_2$, as we illustrate in Fig.~1.

\begin{figure}[h]
\includegraphics[{height=4.0cm,width=9.0cm}]{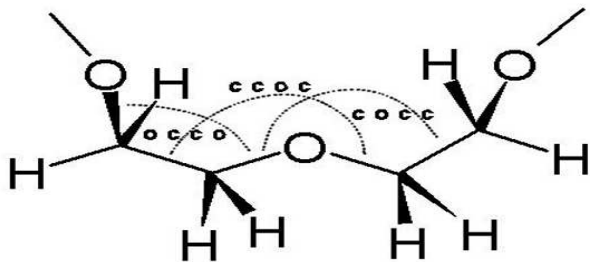}
\caption{The three most important dihedral angles in the POE
main chain.}
\end{figure}

We can visualize topological solitons in the POE chain thinking of the
$tgt$ conformation, the most stable conformational configuration in the POE
chain in liquid water.
The symmetry of the chain admits two equivalent possibilities, where the
helix can be clockwise, generated by a sequence of $tg^{+}t$
arrangements, or counterclockwise, generated by a sequence of
$tg^{-}t$ orientations. These two gauche conformations (helixes) are
degenerate in energy, and they are typically very populated, according
to the previous
discussion \cite{dme-nmr,dme-ir}. They appear to be the most stable of all
the conformational possibilities the POE chain may comprise. Thus, although
the POE chain may adopt other distinct conformations, our
model considers only the two most important set of degrees of freedom,
represented by the O-CH$_2$-CH$_2$-O and CH$_2$-CH$_2$-O-CH$_2$ dihedral
angles, and the most stable conformation, the {\it gauche} and {\it trans}
arrangements, respectively. For both angles, the {\it cis}
conformation is unstable, and is seen as a barrier for {\it gauche} and
{\it trans} orientations to tunnel to its symmetric partner. We illustrate
this in Fig.~2 for the {\it gauche} conformation of the O-CH$_2$-CH$_2$-O
dihedral angle.

\begin{figure}[h]
\includegraphics[height=5.0cm,width=8.5cm]{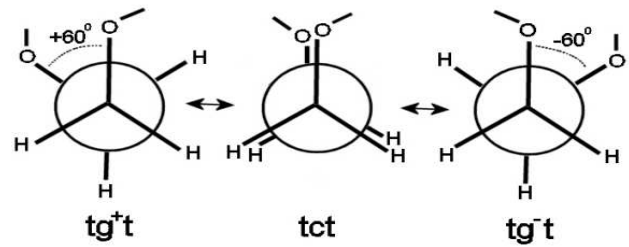}
\caption{The {\it cis} conformation that intermediates the most
stable {\it gauche} conformation for the O-CH$_2$-CH$_2$-O dihedral angle.
The central circles are drawn to represent the two neighboring carbon
atoms between subsequent oxygen atoms.} 
\end{figure}

This reasoning allows introducing a model of two real
scalar fields to describe the nonlinear chain. Under this assumption
we could establish, {\it a priori,} the two most relevant sets of
degrees of freedom to describe the transition
$tg^{+}t\leftrightarrow tg^{-}t$ that springs as the topological soliton.

The connection between these detailed information at molecular level
with the mathematical representation of the polymer is implemented
by the field theoretical model that we now develop. The model
is described by two real scalar fields, $\phi$ and $\chi$,
which live in the bidimensional space-time $(x,t)$. The Lagrangian
density has the form 
\begin{equation}
{\cal L}=\frac{1}{2}\partial_{\alpha }\phi\partial^{\alpha}\phi+
\frac{1}{2}\partial_{\alpha}\chi\partial^{\alpha}\chi-V(\phi,\chi)
\end{equation}
where 
\begin{equation}
V(\phi,\chi)=\frac{1}{2}W_{\phi}^{2}-\frac{1}{2}W_{\chi}^{2}
\end{equation}
Here $\partial_{\alpha}\phi=\partial\phi/\partial x^{\alpha}$,
$W_{\phi}=\partial W/\partial\phi$, etc. We are using standard notation,
with $x^{\alpha}=(x^{0}=t,x^{1}=x)$, and $x_{\alpha}=(x_{0}=t,x_{1}=-x)$,
and $\hbar=c=1$. The form of the potential is special, because it allows
obtaining static solutions of the equations of motion 
\begin{eqnarray}
\frac{d^{2}\phi}{dx^{2}}&=&W_{\phi}W_{\phi\phi}+W_{\chi}W_{\chi\phi}
\\
\frac{d^{2}\chi}{dx^{2}}&=&W_{\phi}W_{\phi\chi}+W_{\chi}W_{\chi\chi}
\end{eqnarray}
through the first order differential equations 
\begin{eqnarray}
\frac{d\phi}{dx}&=&W_{\phi}
\\
\frac{d\chi}{dx}&=&W_{\chi}
\end{eqnarray}
Solutions of the first order equations are stable and are minimum energy
solutions -- see {\cite{bbr00}} and references therein for more details.

The model we consider is defined by the superpotential $W=W(\phi,\chi)$,
which has the specific form 
\be
W=\lambda a^2\phi+\nu a^2\chi-\frac{1}{3}\lambda\phi^3-\mu\phi\chi^2
\end{equation}
We consider $\lambda,\mu,\nu$ real, and $a$ real and positive. This model is
new, and reproduces the model first considered in Ref.~\cite{95} in the
limit $\nu\to0$. We calculate $W_{\phi\phi}$ and $W_{\chi\chi}$ to see that 
\be
\label{w22}
W_{\phi\phi}=-2\lambda\phi,\;\;\;\;\;W_{\chi\chi}=-2\mu\phi
\end{equation}
We notice that the superpotential is harmonic for $\mu+\lambda=0$,
thus the equations of motion can be reduced to a family of first
order equations for $\lambda+\mu=0$ \cite{bms01}.

The potential presents absolute minima at the points
$({\bar \phi}_i,{\bar\chi}_i)$, where
${\bar\chi}_i/a=\nu a/2\mu{\bar\phi}_i$ are given in terms of
${\bar\phi}_i$, which represent the real solutions of 
\be
\left(\frac{\bar\phi}{a}\right)^2=\frac{1}{2}\pm\frac{1}{2}
\sqrt{1-\frac{\nu^2}{\mu\lambda}\,}
\end{equation}
There are two solutions for $\nu^2/\lambda\mu<0$, four solutions for
$0<\nu^2/\mu\lambda<1$, two solutions for $\nu^2/\lambda\mu=1$, and no
solution for $\nu^2/\mu\lambda>1$. We consider the case
$0<\nu^2/\mu\lambda<1$. We write $\mu=\lambda r$, and $\nu=\lambda s$,
to get to $0<s^2/r<1$. This implies that $r>0$, and we take
$-\sqrt{r}<s<\sqrt{r}$. 

The equations of motion for static fields are solved by solutions to the
first order differential equations, after changing $\phi\to a\phi$, $\chi\to
a\chi$, and $x\to\lambda a x$ 
\begin{eqnarray}
\frac{d\phi}{dx}&=&1-\phi^2-r\chi^2 \\
\frac{d\chi}{dx}&=&s-2r\phi\chi
\end{eqnarray}
We notice that this system does not allow one-field solutions for $s\neq0$.
This peculiar behavior of the model is important, and is in good
agreement with the quantum chemistry investigation that we have implemented,
as we comment on below. We can find true two-field solutions
to the above first order equations. They are described by the field
configurations
\begin{eqnarray} 
\label{sol1}
\phi_s(x)&=&a\sqrt{\frac{1}{2r}\,}\,\tanh(\lambda a\sqrt{2r}\,x)
\\
\chi_s(x)&=&\pm a\sqrt{\frac{1}{r}-
\frac{1}{2r^2}\,}\,\tanh(\lambda a\sqrt{2r}\,x)
\label{sol2}
\end{eqnarray}
which are valid for $s=\pm\sqrt{2-1/r}$, with $r>1/2$. We notice that the
limit $r\to1/2$ transforms the above solution back to the one-field solution
of the simpler model, where $s=0$. We see that the amplitude of $\phi$ is
$A_{\phi}=a/\sqrt{2r}$, and that of $\chi$ reads $A_{\chi}=|s\,|\,A_{\phi}$.
Thus, for $1/2<r<1$, we get $0<s^2<1$, and the amplitude of $\chi$ is always
smaller then that of $\phi$. For $r>1$ we get $s^2>1$, and in this case the
amplitude of $\chi$ is greater then that of $\phi$. A complete investigation
of the two-field model will be published elsewhere. 

In the POE polymer, the topological soliton represents a conformational
distortion in the $tgt$ chain: it describes the change from the $tg^+t$
conformation to the $tg^-t$ conformation. This modification is done with the
chain passing through the unstable {\it cis} conformations, in a way such that
the topological distortions locally increase the dipole moment in the chain,
which are evenly distributed along the entire chain in both
the $tg^+t$ and $tg^-t$ conformations.
The presence of a topological soliton breaks the dipole moment uniformity
in the chain, increasing the polar nature of the chain in the soliton core,
and this is the mechanism that triggers the process for capturing
metallic ions in aqueous solution, or that favors the presence of hidrogen
bonds in water solutions.

To make our investigations more quantitative we now examine
the connections between the detailed informations obtained in the model
described by two real scalar fields and the POE chain at molecular level.
This can be obtained through quantum chemistry calculations, with the
mathematical representation of the polymer as given by the field theoretical
model. We investigate the chain H(-CH$_2$-CH$_2$-O-)$_{40}$H, a POE chain
with 40 monomeric units. We suppose that the topological soliton
represents the transition state between the positive and
negative {\it helix} conformations for the polymer main chain.
The transition state between the two most stable conformations of the POE
is represented by a polymer that mixes the {\it gauche}
and the energetically unfavorable {\it cis} conformations at the same
molecule, in a way such that it makes possible the application of quantum
chemistry machinery to do the investigation, since the soliton excitation
is a continuum phenomenon that needs an extended molecular representation.
In Fig.~3 one depicts the polymer molecules considered as reagent,
named positive helix (lower figure), and as transition state, named soliton
(upper figure). We established that the dihedral
angle C-O-C-C was maintained at configuration trans ($\pi$) throughout
the entire polymer molecule, and that the other two dihedral
angles in the main chain -- namely O-CH$_2$-CH$_2$-O and
CH$_2$-CH$_2$-O-CH$_2$ -- are choosen according to the field theoretical
model, in which we are identifying the $\phi$ and $\chi$
fields with the two dihedral angles CH$_2$-CH$_2$-O-CH$_2$ and
O-CH$_2$-CH$_2$-O, respectively.
[In fact, we have done several other investigations: for instance, if the
CH$_2$-CH$_2$-O-CH$_2$ angle is also frozen, the results are very unusual,
showing that the polymer main chain visit very energetic configurations,
returning along the same chain axis, but reversing its direction; also, if
all the three dihedral angles are free, the polymer main chain bends
considerably, breaking the unidirecional character of the polymer].

\begin{figure*}
\includegraphics[height=3.0cm,width=18.0cm]{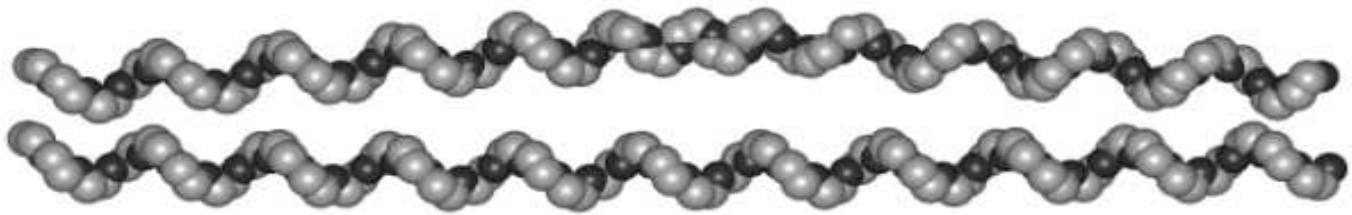}
\caption{The two conformational structures for the POE main chain.
The lower chain is shown in the positive helix conformation. The upper
chain shows the solitonic structure, which drives the most important dihedral
angles in the POE main chain, changing the {\it gauche} conformation from
positive (negative) to negative (positive) helicity.}
\end{figure*}

We use results of the two field model to see that the modification of the
two dihedral angles that are correlated with important degrees of freedom
of the main chain follow the hyperbolic tangent behavior. We identify the
two dihedral configurations O-CH$_2$-CH$_2$-O and CH$_2$-CH$_2$-O-CH$_2$
with the $\chi$ and $\phi$ fields, respectively, and so we use the
quantitative pattern of the soliton
solution (\ref{sol1}) and (\ref{sol2}) to set the values for dihedral angles
through the 40 monomeric unities of the POE molecule. The positive helicity
is achieved by setting the O-CH$_2$-CH$_2$-O and CH$_2$-CH$_2$-O-CH$_2$
angles with the values $+\pi/3$ and $-\pi$, respectively, in the
entire polymer molecule -- see Fig.~3, lower chain. For the soliton,
however, we consider the first $8$ monomeric units in both extremities
of the polymer as positive and negative
helicities, respectively, while for the 24 central units the values of the
O-CH$_2$-CH$_2$-O and CH$_2$-CH$_2$-O-CH$_2$ dihedral angles vary according
to the hyperbolic tangent profile. The O-CH$_2$-CH$_2$-O angle was modified
from $+\pi/3$ to $-\pi/3$, while the CH$_2$-CH$_2$-O-CH$_2$ angles was
modified from $-\pi$ to $+\pi$, through the curvature of each one of the
two hyperbolic tangent solutions that describe the soliton. This
identification restrains the parameters $r$, $s$, and $a$ that appear
in the field theoretical model to the values $r=9/17$, $s=-1/3$, and
$a=\pi\,\sqrt{18/17}$ and the soliton solution can be rewritten as
\ben
\phi(x)&=&\pi\,\tanh\left(\sqrt{18/17\,}\; x\right)
\\
\chi(x)&=&-(\pi/3)\,\tanh\left(\sqrt{18/17\,}\; x\right)
\een
where we have changed $\lambda a x\to x$, to use dimensionless coordinate.

The molecules shown in Fig.~3 represent the final results
obtained with semi-empirical molecular orbital calculations that were made
by using AM1 \cite{AM1} method implemented in the MOPAC 93 version 6.0
package \cite{mopac1}. During the geometry optimization procedures used in
order to find the stationary point at the potential energy surface (PES) in
each case (positive helicity and soliton), all the dihedral angles
were maintained frozen, while the normal angles and bond lengths were
allowed to relax. In Fig.~3 we notice that the upper chain (which contains
the soliton) slightly deviates from its unidirensional character, but
the linear configuration is still in good approximation with the field model.
The main results for the electronic properties are
given in the Table below, which shows the heat of formation and
dipole moment ($\mu$) for both the soliton (upper line)
and the helix (lower line) molecules.

\begin{center}
\begin{tabular}{|c|c|}
\hline
{\,Heat of Fomation (kJ)} & {\,$\mu$ (Debye)} \\ \hline\hline
{-7621.47} & {11.53} \\ \hline
{-7719.56} & {1.42} \\ \hline
\end{tabular}
\end{center}

The results reveal that the soliton is 98.09 kJ more energetic
than the normal {\it gauche} conformation, in agreement with the fact
that the soliton induces a localized transition between positive and
negative helicities of the {\it gauche} conformation.
This energy difference is explained by the instability of the {\it cis}
or near-{\it cis} conformations in both O-CH$_2$-CH$_2$-O and
CH$_2$-CH$_2$-O-CH$_2$ dihedral angles, found in the soliton core.
We believe that it can be minimized in condensed media,
particularly in polar solvents like water, for example, because of the
greater polarity of the soliton molecule ($\mu =11.53$ D) in comparison
to the positive or negative helicity ($\mu=1.42$ D).

The soliton model that we have used seems to be limited, since it does
not account for intermolecular interactions between neigbour POE chains.
However, this is a good approximation in aqueous solutions, even at high
concentrations, since interactions between POE chains via hydrogen bonds are
scarce \cite{prl2000}. In fact, there are more interactions between
oxyethylene units in the same chain \cite{prl2000}, mediated by
hydrogen bonds via neighbour water molecules, but this we have empirically
taken into account by considering the positive and negative helix
conformations for the POE main chain as the most populated conformations
in aqueous solutions.

We thank A.R. Gomes, R. Longo, and E. Ventura for stimulating discussions,
and CAPES, CNPq, PROCAD and PRONEX for partial support.


\begin{references}
\bibitem{ssh79}W.-P. Su, J.R. Schrieffer, and A.J. Heeger, Phys. Rev.
Lett. {\bf42}, 1698 (1979).
\bibitem{hks88}A.J. Heeger, S. Kivelson, J.R. Schrieffer, and W.-P. Su,
Rev. Mod. Phys. {\bf60}, 781 (1988).
\bibitem{jsc81}R. Jackiw and J.R. Schrieffer, Nucl. Phys. B {\bf190}
[SF3], 253 (1981).
\bibitem{jac95}R. Jackiw, {\it Diverse Topics in Theoretical and
Mathematical Physics} (World Scientific, Singapore, 1995).
\bibitem{mbo78}M.L. Mansfield and R.H. Boyd, J. Polym. Sci. Phys. Ed.
{\bf16}, 1227 (1978).
\bibitem{80}M.L. Mansfield, Chem. Phys. Lett. {\bf 69}, 383 (1980); J.L.
Skinner and P.G. Wolynes, J. Chem. Phys. {\bf 73}, 4015 and 4022 (1980);
J.L. Skinner and Y.H. Park, Macromolecules {\bf17}, 1735 (1984).
\bibitem{zco93}F. Zhang and M.A. Collins, Chem. Phys. Lett. {\bf214}, 459
(1993); Phys. Rev. E {\bf49}, 5804 (1994).
\bibitem{we}D. Bazeia and E. Ventura, Chem. Phys. Lett. {\bf303}, 341 (1999);
E. Ventura, A.M. Simas, and D. Bazeia, Chem. Phys. Lett. {\bf320}, 587 (2000).
\bibitem{95}  D. Bazeia, M.J. dos Santos, and R.F. Ribeiro, Phys. Lett. A 
{\bf208}, 84 (1995); D. Bazeia, R.F. Ribeiro, and M.M. Santos, Phys. Rev.
D {\bf54}, 1852 (1996); D. Bazeia, R.F. Ribeiro, and M.M. Santos, Phys. Rev.
E {\bf54}, 2943 (1996).
\bibitem{bnr}D. Bazeia, J.R.S. Nascimento, R.F. Ribeiro, and D. Toledo, J.
Phys. A {\bf30}, 8157 (1997); C.A.G. Almeida, D. Bazeia, and L. Losano,
J. Phys. A {\bf34}, 3351 (2001).
\bibitem{bbr00}D. Bazeia and F.A. Brito, Phys. Rev. Lett. {\bf84}, 1094
(2000); Phys. Rev. D {\bf61}, 105019 (2000).
\bibitem{ilg}A. Alonso Izquierdo, M.A. Gonz\'alez Le\'on, and J. Mateos
Guilarte, Phys. Lett. B {\bf480}, 373 (2000); Phys. Rev. D {\bf65},
085012 (2002). 
\bibitem{har92}J.M. Harris, Ed. {\it Poly(ethylene glycol) Chemistry:
Biotechnical and Biomedical Applications} (Plenum, New York, 1992).
\bibitem{jpc93}G.D. Smith, R.L. Jaffe, and D.Y. Yoon,
J. Phys. Chem. {\bf97}, 12752 (1993).
\bibitem{jacs1995}G.D. Smith, R.L. Jaffe, and D.Y. Yoon,
J. Am. Chem. Soc., {\bf117}, 530 (1995).
\bibitem{jpca97}G.D. Smith, R.L. Jaffe, and H. Partridge, J. Phys. Chem.
A {\bf101}, 1705 (1997).
\bibitem{tas96}K. Tasaki, J. Am. Chem. Soc. {\bf118}, 8459 (1996).
\bibitem{tas99}K. Tasaki, Comp. Theo. Polym. Sci. {\bf9}, 271 (1999).
\bibitem{peg01}J.-P. Wan et al. J. Controlled Release {\bf75}, 129 (2001).
\bibitem{mat02}S.A. Wahab and H. Matsuura, J. Mol. Struct. {\bf606}, 35 (2002).
\bibitem{smi96}G.D. Smith et al. Macromolecules, {\bf29}, 3462 (1996).
\bibitem{prl2000}G.D. Smith, D. Bedrov, and O. Borodin, Phys. Rev. Lett.
{\bf85}, 5583 (2000).
\bibitem{poe-raman}H. Matsuura and K. Fukuhara, J. Mol. Structure,
{\bf126}, 251 (1985).
\bibitem{dme-nmr}K. Inomata and A. Abe, J. Phys. Chem. {\bf96}, 7934 (1992).
\bibitem{dme-ir}H. Yoshida et al. Chem. Phys. Lett. {\bf196}, 601 (1992).
\bibitem{kf81}R. Kjellander and E. Florin, J. Chem. Soc., Faraday
Trans. 1 {\bf77}, 2053 (1981). 
\bibitem{ngm01}Z.S. Nickolov, N. Goutev, and H. Matsuura, J. Phys. Chem. A
{\bf105}, 10884 (2001).
\bibitem{bms01}D. Bazeia, J. Menezes, and M.M. Santos,
Phys. Lett. B {\bf521}, 418 (2001); Nucl. Phys. B {\bf636}, 132 (2002).
\bibitem{AM1}M.J.S. Dewar et al. J. Am. Chem. Soc.,{\bf107}, 3902 (1985).
\bibitem{mopac1}J.J.P. Stewart. J. Comput.-Aided Mol. Des. {\bf4}, 1
(1990).
\end{references}
\end{document}